\documentclass[a4paper,11pt]{article}
\usepackage{makeidx}
\usepackage{amsmath}
\usepackage{latexsym}
\usepackage{amssymb}

\usepackage{graphicx}
\usepackage{float}
\begin{document}

\hbox to 16.5 true cm{\hfill TOKAI-HEP/TH-04-09\ \ \ \ \ \ \ }

\begin{center}
{\Large\bf General Relativity with Torsion}
\\
\vspace{16pt}
Tomoki Watanabe${}^*$ and Mitsuo J. Hayashi${}^\dag$
\\
\vspace{16pt}
{\it\footnotesize Department of Physics, Tokai University, 1117, Kitakaname, Hiratsuka, 259-1292, Japan}

{\footnotesize
${}^*$2aspd004@keyaki.cc.u-tokai.ac.jp;
${}^\dag$mhayashi@keyaki.cc.u-tokai.ac.jp}
\\

\vspace{12pt}
\end{center}
\begin{abstract}
General Relativity with nonvanishing torsion has been investigated in the first order formalism of Poincar\'e gauge field theory.
In the presence of torsion, either side of the Einstein equation has the nonvanishing covariant divergence.
This fact turned out to be self-consistent in the framework under consideration.
By using Noether's procedure with the definition of the Lie derivative where the general coordinate transformation and the local Lorentz rotation are combined, the revised covariant divergence of the energy momentum is consistently obtained.
Subsequently we have definitely derived the spin correction to the energy momentum tensor for the Dirac field and the Rarita-Schwinger field in the Einstein equation. We conjecture that the accelerated expansion of the universe possibly arises due to the spin correction in our framework.
\end{abstract}
\begin{center}
\end{center}

\section{Introduction}

In the well-known formulation of General Relativity, gravity is appropriately described by Riemann geometry which is torsion-free. The metric of the space-time represents the gravitational field.
The connections are given by the Christoffel symbol compatible with the metric structure.
However in particle physics, the spinor fields are the essential ingredients in the description of natural law. Therefore, in order to bring spinors into the curved space-time, the vierbein (tetrad) had been introduced to formulate General Relativity by the local gauge transformation invariance, such as the local translation or the local Poincar\'e group\cite{hehl}-\cite{shapiro}. 

If one incorporates spinors as a member of physical objects into gravitational theory, the assumption of torsion-free seems unnatural.
Then, one should introduce Cartan geometry which is characterized by torsion as well as curvature and which provides the field equation relating torsion to spin in the Poincar\'e gauge theory\cite{utiyama,hayashi}.
For example, the theory of supergravity which includes bosons together with fermions inevitably possesses torsion as the essential part of the theory\cite{nieuwen}-\cite{weinberg}.

With or without supersymmetry, one of the advantages of introducing torsion is to modify existing cosmology based on usual General Relativity by dint of the intrinsic spin of matter.
In pioneering works\cite{trautmann}-\cite{kuchowicz}, it was shown that torsion may remove the initial singularity.
Subsequently, torsion was used to investigate the effects of spinning matter in the early universe\cite{nurgaliev}.
Moreover, it turned out that the spin-dominated universe causes inflation in the context of Cartan geometry\cite{gasperini,kim}.
Recently, Szydlowski and Krawiec studied the possibility that the cosmological model with spinning matter might be a candidate to explain the accelerated expansion of the present universe\cite{szydlowski}, and concluded that the spinning matter is not an alternative to the cosmological constant but is admissible.
In such cosmology, it is important to note that the ``spinning fluid" filling in the universe is constructed under the assumption that thermodynamical variables can be introduced\cite{ray}.

In this paper, having a perspective to establish a new approach to cosmology, we investigate the modifications caused by torsion in General Relativity by revisiting the Poincar{\'e} gauge field theory of gravity without the supersymmetry.
Starting from the first order formalism, where the vierbein and the spin connection are independent fields, we derive the revised conservation law of the energy momentum tensor with torsion by constituting an alternative Lie derivative including the local Lorentz transformation, which seems complementary to the method in Ref.\cite{hehl}.
Because the torsion and the spin density of matter fields are convertible, a part of the Einstein tensor originated from torsion can be regarded as correction to the energy momentum tensor due to the spin density.  
In particular, we definitely derive the correction to the energy momentum tensor for the Dirac field and the Rarita-Schwinger field which is expressed in terms of products of the spin density, and prove the accelerated expansion of the universe can arise without introducing the thermodynamical variables.

This paper is orgnized as follows. In section 2, in order to give some definitions, we briefly review the vierbein formalism\cite{weinberg,aldrovandi} which is convenient to describe the couplings of gravity to spinor fields\cite{penrose, birrell}. It will turn out that the vierbein formalism produces an extension of usual Einstein's theory. We turn section 3 to derive  explicitly the revised formulae due to the torsion \cite{hehl,hehl2,shapiro,utiyama,penrose}. Subsequently, the concrete incorporation of the Dirac and Rarita-Schwinger fields into General Relativity will be demonstrated. Finally, in section 4 we present our conclusions.

\section{Vierbein formalism}

\subsection{Vierbein postulate}

The metric tensor $g_{\mu\nu}$ adequately describes gravitational theory which contains matter fields consist of scalar, vector, and tensor fields. Vectors in this theory carry linear representation of $GL(4,{\bf R})$, if the connection $\Gamma$ which belongs to $gl(4,{\bf R})$ is given.

Let us demand that a scalar product should be invariant under the parallel transport, i.e., the metricity condition
\begin{equation}
\nabla_\rho g_{\mu\nu}= 0,
\label{metr.cond.}
\end{equation}
where the covariant derivative is defined by using the connection $\Gamma^\rho_{\mu\nu}$, which is not necessarily symmetric, as $\nabla_\nu V^\mu\equiv \partial_\nu V^\mu +\Gamma^\mu_{\rho\nu}V^\rho$. We will not deal with non-metricity here. We assume that one can always take the Minkowski metric $\eta_{ab}={\rm diag}(+,-,-,-)$ at any space-time point. In a word, the world distance requires to be expressed in two ways:
$
ds^2=g_{\mu\nu}dx^\mu dx^\nu=\eta_{ab}dX^a dX^b.
$
One can choose an orthonormal frame $e_a^{\ \mu}$ there, which transforms as a vector under both $GL(4,{\bf R})$ and $SO(3,1)$ according to each index.
Then, $e_a^{\ \mu}$ gets more essential than $g_{\mu\nu}$; the metric can be expressed by a square of $e_a^{\ \mu}$:
\begin{equation}
g_{\mu\nu}=\eta_{ab}e_\mu^{\ a}e_{\nu}^{\ b},
\label{e square}
\end{equation}
which behaves as a $GL(4,{\bf R})$ (world) tensor and a $SO(3,1)$ (Lorentz) scalar. In general, $e_a^{\ \mu}$ and $e_\mu^{\ a}$ project world vectors $V^\mu$ onto Lorentz components $V^a$ and vice versa, e.g., $V^a=e_\mu^{\ a}V^\mu$.

Note that $g_{\mu\nu}$ has 10 components while $e_a^{\ \mu}$ has 16 components. In order that the world distance in two ways be identified, the defference 6 components must be compensated; the Lorentz transformation $\Lambda^a_{\ b}$ plays this role. $e_a^{\ \mu}$ is called vierbein or tetrad.

Let $\omega^{ab}_\mu$ be the connection for the orthonormal frame. The metricity condition (\ref{metr.cond.}) in terms of $\eta_{ab}$ and $\omega^{ab}_\mu$ reduces to the Lorentz algebra $so(3,1)$. Therefore, by using $\omega^{ab}_\mu$, we can define the covariant derivative for any field $\phi$ which transforms under the irreducible representation of $SO(3,1)$:
$
D_\mu\phi^A\equiv\partial_\mu\phi^A+(\Omega_\mu)^A_{\ B}\phi^B,
$
where the index $A$ means a general type or components of fields and
$
\Omega_\mu\equiv (i/2)\omega^{ab}_\mu S_{ab}
$
is an element of $so(3,1)$.
For example, $D_\mu$ acts on a spinor as
\begin{equation}
D_\mu\psi=\partial_\mu\psi-\frac{i}{4}\omega^{ab}_\mu\sigma_{ab}\psi,
\label{cov.spinor}
\end{equation}
since the generator $S_{ab}$ for spinors is given by
$S_{ab}=-\sigma_{ab}/2\equiv -(i/4)[\gamma_a,\gamma_b]$.
Here $\gamma_a$ indicates the $\gamma$-matrices which always take constant values.
Generically, under the local Lorentz transformation $D[\Lambda(x)]$, $D_\mu\phi$ covariantly transforms as $D_\mu\phi\rightarrow D[\Lambda(x)](D_\mu\phi)$, while $\Omega_\mu$ obeys an inhomogeneous transformation law; the infinitesimal representation $D[\Lambda(x)]=1+(i/2)\theta^{ab}(x)S_{ab}$ leads to
\begin{equation}
\omega^{ab}_\mu\rightarrow
\omega^{ab}_\mu+\theta^a_{\ c}\omega^{cb}_\mu+\theta^b_{\ c}\omega^{ac}_\mu
-\partial_\mu\theta^{ab}.
\label{inf.tr.}
\end{equation}

Also note that 40 constraints of the metricity condition (\ref{metr.cond.}) balance the difference between 24 components of $\omega^{ab}_\mu$ and 64 components of $\Gamma^\rho_{\mu\nu}$. $\omega^{ab}_\mu$ (or $\Omega_\mu$) is sometimes called spin connection.

If, as more essential quantity, the vierbein is employed to describe gravity instead of the metric, then one may wish to establish a more fundamental condition for the vierbein to induce the metricity condition (\ref{metr.cond.}). Since $e_a^{\ \mu}$ has both world and Lorentz indices, we should replace $\nabla_\mu$ in Eq.(\ref{metr.cond.}) by $\mathcal{D}_\mu$ which operates on these indices:
\begin{equation}
\mathcal{D}_\nu e^{a\mu}\equiv
\partial_\nu e^{a\mu} +\Gamma^\mu_{\rho\nu}e^{a\rho}+\omega^{ab}_\nu e_b^{\ \mu}= 0.
\label{vierb.cond.}
\end{equation}
We call the formula (\ref{vierb.cond.}) as ``vierbein postulate". This postulate seems to determine a form of the spin connection, i.e.,
$
\omega^{ab}_\mu =g^{\rho\nu}e_\rho^{\ a}\nabla_\mu e_\nu^{\ b}.
$
Actually, this form is however not unique to satisfy the above transformation law. One can so divide $\omega^{ab}_\mu$ into two parts that one of them follows the law and another transforms as a Lorentz tensor and besides a world vector (see Eq.(\ref{omega}) and Eq.(\ref{omega2})). This means that there is an important freedom on the choice of $\omega^{ab}_\mu$ when one demands the Lorentz covariance of the theory.

\subsection{Curvature and torsion}

Now we can define curvature. The field strength
$F_{\mu\nu}\equiv [D_\mu, D_\nu]$
also belongs to $so(3,1)$; curvature tensor $R^{ab}_{\ \ \mu\nu}$ is defined to be its coefficients: $F_{\mu\nu}=(i/2)R^{ab}_{\ \ \mu\nu}S_{ab}.$
From the definition of $D_\mu$ and $so(3,1)$ algebra, we have
\begin{equation}
R^{ab}_{\ \ \mu\nu}=
\partial_\mu\omega^{ab}_\nu+\omega^{ac}_\mu\omega^{\ \ \> b}_{\nu c}
-\partial_\nu\omega^{ab}_\mu-\omega^{ac}_\nu\omega^{\ \ \> b}_{\mu c}.
\label{curvature omega}
\end{equation}
By definition, it follows that $R^{ab}_{\ \ \mu\nu}$ transforms as a world and Lorentz tensor of second rank respectively. We can therefore construct a world tensor of fourth rank by projecting on the world components,
$
R^\rho_{\ \sigma\mu\nu}=e_a^{\ \rho}e_{b\sigma}R^{ab}_{\ \ \mu\nu},
$
which is just equal to the form of usual Riemann curvature
\begin{equation}
R^\rho_{\ \sigma\mu\nu}=
\partial_\mu\Gamma^\rho_{\sigma\nu}+\Gamma^\rho_{\lambda\mu}\Gamma^\lambda_{\sigma\nu}
-\partial_\nu\Gamma^\rho_{\sigma\mu}-\Gamma^\rho_{\lambda\nu}\Gamma^\lambda_{\sigma\mu},
\label{curvature}
\end{equation}
together with the vierbein postulate (\ref{vierb.cond.}),
$
\Gamma^\rho_{\mu\nu}=e_a^{\ \rho}D_\nu e_\mu^{\ a}.
$
Thus, it is conceivable that $\omega^{ab}_\mu$ may also be more essential than $\Gamma^\rho_{\mu\nu}$. If once $\omega^{ab}_\mu$ is given, the covariant derivative for any spinor (\ref{cov.spinor}) can be defined. Because spinor is a world scalar, its action with Eq.(\ref{cov.spinor}) is just desired one, i.e., we can form gravitational theory which includes spinor fields by starting from $e_a^{\ \mu}$, $\omega^{ab}_\mu$ and the vierbein postulate.

Note that the vierbein postulate (\ref{vierb.cond.}) suggests $\Gamma^\rho_{\mu\nu}$ is not necessarily symmetric. An antisymmetric part of $\Gamma^\rho_{\mu\nu}$ which is a world tensor is called torsion:
\begin{equation}
C^{\rho}_{\;\;\mu\nu}\equiv\Gamma^\rho_{\mu\nu}-\Gamma^\rho_{\nu\mu}
\equiv 2\Gamma^\rho_{[\mu\nu]}.
\label{torsion}
\end{equation}
If torsion is nonzero, the Einstein tensor $G_{\mu\nu}$ constructed by the curvature (\ref{curvature}) becomes asymmetric. Consequently, the energy momentum tensor of matter $T_{\mu\nu}$, which is usually defined as a symmetric tensor, has to be redefined; we will demonstrate the redefinition in the next subsection.

Torsion (\ref{torsion}) can be written as $C=C(e,\partial e,\omega)$ in combination with the vierbein postulate (\ref{vierb.cond.}).
Inversely, $\omega=\omega(e,\partial e,C)$ can be found; using Eq.(\ref{metr.cond.}) and Eq.(\ref{torsion}), one can split up $\Gamma^\rho_{\mu\nu}$ into
\begin{equation}
\Gamma^\rho_{\mu\nu}=
{\rho \atopwithdelims\{\} \mu\nu}+K^{\rho}_{\;\;\mu\nu},
\label{gamma}
\end{equation}
where ${\rho\atopwithdelims\{\}\mu\nu} ={\rho\atopwithdelims\{\}\nu\mu}$ is Christoffel symbol and $K^{\rho}_{\;\;\mu\nu}$ is called contortion, given by $C^{\rho}_{\;\;\mu\nu}$:
$
K^{\rho}_{\;\;\mu\nu}=
(C^{\rho}_{\;\;\mu\nu}+C^{\ \ \>\rho}_{\mu\nu}+C^{\ \ \>\rho}_{\nu\mu})/2.
$
We find from Eq.(\ref{gamma}) a similar expression:
\begin{equation}
\omega^{ab}_\mu=\tilde{\omega}^{ab}_\mu +K^{ab}_{\ \ \mu},
\label{omega}
\end{equation}
where $\tilde{\omega}^{ab}_\mu$ depends on only $e_a^{\ \mu}$
\footnote{We will use a tilde to indicate a quantity irrelevant to torsion throughout this paper.}:
\begin{equation}
\tilde{\omega}^{ab}_\mu=
e^{a\rho}\partial_{[\mu} e_{\rho]}^{\ b}
+\frac{1}{2}e^{a\rho}e^{b\sigma}e_{c\mu}\partial_{[\sigma} e_{\rho]}^{\ c}
-(a\leftrightarrow b),
\label{omega2}
\end{equation}
and $K^{ab}_{\ \ \mu}\equiv e_\rho^{\ a}e^{b\mu}K^{\rho}_{\;\;\mu\nu}$ is Lorentz-tensorial as well as world-vectorial since $K^{\rho}_{\;\;\mu\nu}$ is a pure world tensor. Clearly, $K^{ab}_{\ \ \mu}$ does not contribute to the Lorentz transformation law (\ref{omega2}) of $\Omega_\mu$. In fact a differentiation of $e_a^{\ \mu}$ in Eq.(\ref{omega2}) just causes the inhomogeneous term in the transformation law (\ref{inf.tr.}). In other words, the Lorentz covariance of the theory can be realized if only $\tilde{\omega}^{ab}_\mu$ is introduced as the connection. Thus one can choose $\omega^{ab}_\mu$ so that torsion vanishes in gravitational theory which contains spinor fields. However, we leave torsion nonvanishing; our main purpose is to view the modifications due to torsion in General Relativity.

\subsection{Field equations}

Let us formulate the field equations which gravity and torsion follow in the vierbein formalism\cite{hehl}-\cite{shapiro}.
For this purpose, it is convenient to take the first order (Palatini) formalism\cite{nieuwen,shapiro,buch}.

In usual General Relativity, this method is to regard the metric and the symmetric connection as independent. Indeed, without the metricity condition, the variation of the Einstein-Hilbert action $S_{\rm g}(g,\Gamma,\partial\Gamma)$ with respect to $\Gamma^\rho_{\mu\nu}$ gives the unique expression of the Christoffel symbol as a field equation $\Gamma=\Gamma(g,\partial g)$.
In other words, the metricity condition can be obtained as the field equation, and then the first order and the second order formalism are equivalent on the classical level.

If the connection is asymmetric, then one has to take into account the additional independent variable $K^{\rho}_{\;\;\mu\nu}$ which is so far unknown.
However, by substituting the metricity condition as the field equation into the variation of $S_{\rm g}$ with respect to $K^{\rho}_{\;\;\mu\nu}$, the desired equation which torsion follows can be found\cite{hehl}. 
Therefore, alternatively starting with the independent variables $g_{\mu\nu}$, asymmetric
$\Gamma^\rho_{\mu\nu}={\rho \atopwithdelims\{\} \mu\nu}+K^{\rho}_{\;\;\mu\nu}$,
and the metricity condition as a demand, we can also obtain the field equation for torsion.

We have now independent variables $e_a^{\ \mu}$ and $\omega^{ab}_\mu$ in the Einstein-Hilbert Lagrangian:
\begin{equation}
\mathfrak{L}_{\rm g}(e,\omega,\partial\omega )=
-\frac{e}{16\pi G}e_a^{\ \mu}e_b^{\ \nu}R^{ab}_{\ \ \mu\nu}
(\omega,\partial\omega),
\label{eh lagrangian}
\end{equation}
where $e\equiv\det(e_a^{\ \mu})$ and $G$ is the gravitational constant.
Because Eq.(\ref{omega}) has been obtained, one can expect to get the field equation for torsion by imposing the vierbein postulate on the variation of the Lagrangian (\ref{eh lagrangian}) with respect to $\omega^{ab}_\mu$ on the analogy of the above-mentioned scheme.

On the other hand, the Lagrangian of matter fields has the independent variables $\phi$ and $\partial\phi$ additionally.
Suppose that the matter fields minimally couple to gravity, i.e., the $\omega$-dependence of the matter Lagrangian appears only through the covariant derivative $D_\mu\phi$. And if the Lagrangian contains the world covariant derivative operator, we opt to convert them into the Lorentz covariant derivative, e.g., $\nabla_\mu V^\nu=e_a^{\ \nu}D_\mu V^a$. The vierbein postulate (\ref{vierb.cond.}) guarantees this conversion to be always possible. Thus we can set the independent variables in the matter Lagrangian as
\begin{equation}
\mathfrak{L}_{\rm m}=
\mathfrak{L}_{\rm m}(e,\omega,\phi,\partial\phi)
=\mathfrak{L}_{\rm m}(e,\phi,D\phi).
\label{matt.lagrangian}
\end{equation}
Then, the equation of motion for $\phi$ is readily derived:
\begin{equation}
\frac{\partial\mathcal{L}_{\rm m}}{\partial\phi^A}
-(\mathcal{D}_\mu-C_\mu)
\frac{\partial\mathcal{L}_{\rm m}}{\partial D_\mu\phi^A}
+\Gamma^\rho_{\nu\mu}(s^\nu_{\ \rho})_A^{\ \>B}
\frac{\partial\mathcal{L}_{\rm m}}{\partial D_\mu\phi^B}
=0,
\label{eom}
\end{equation}
where $\mathcal{L}_{\rm m}\equiv\mathfrak{L}_{\rm m}/e$,
$C_\mu\equiv C^{\nu}_{\;\;\mu\nu}$,
and $s^\nu_{\ \rho}$ is the representation matrix of the infinitesimal general coordinate transformation appropriate to $\phi$. The last term in Eq.(\ref{eom}) becomes significant only when the index $A$ carries world components; Rarita-Schwinger field is a candidate for such a field.

Although, of cause, the usual variational method (which $\phi$ and $\partial\phi$ are regarded as independent) also yields the same equation of motion, the concrete calculation is typically tedious. The merit of taking variable $D_\mu\phi$ as independent is to offer an easier method of the calculation to find the equation of motion.

Now, as is well-known, the variation of the total Lagrangian
$\mathfrak{L}=\mathfrak{L}_{\rm g}+\mathfrak{L}_{\rm m}$
with respect to $e_a^{\ \mu}$ gives the Einstein equation in the vierbein formalism:
\begin{equation}
G_\mu^{\;\; a}\equiv
R_\mu^{\;\; a}-\frac{1}{2}e_\mu^{\ a}R
=8\pi G\>T_{\mu}^{\ a},
\end{equation}
with the Ricci tensor
$R_\mu^{\;\; a}\equiv e_b^{\ \nu}R^{ba}_{\ \ \nu\mu}$
and the Ricci scalar
$R\equiv e_a^{\ \mu}e_b^{\ \nu}R^{ab}_{\ \ \mu\nu}$.
For the Lagrangian of matter, we will present specific forms in the section 3.2.

Let us define world tensors as
$G_{\mu\nu}\equiv e_{a\nu}G_{\mu}^{\;\;a}$,
$T_{\mu\nu}\equiv e_{a\nu}T_{\mu}^{\ a}$. It is obvious that these are not symmetric, i.e., one should distinguish $G_{\mu}^{\;\;a}$ from $G^a_{\ \mu}$ etc. The latter means in our notation $G^a_{\ \mu}=e^{a\nu}e_{b\mu}G_{\nu}^{\;\;b}$. The asymmetric Einstein equation is found in this way:
\begin{equation}
G_{\mu\nu}=8\pi G\>T_{\mu\nu}.
\label{E eq2}
\end{equation}

Next, we consider the variation of the action with respect to $\omega^{ab}_\mu$. As mentioned above, using the vierbein postulate after taking the variation of Eq.(\ref{eh lagrangian}), we find
\begin{equation}
-\frac{1}{16\pi G}
\left(
C^{\mu}_{\;\;ab}+2e_{[a}^{\ \mu}C_{b]}
\right)
=
\frac{i}{2}
\frac{\partial\mathcal{L}_{\rm m}}{\partial D_\mu\phi^A}
(S_{ab})^A_{\ B}\phi^B
\equiv
S^\mu_{\;\;ab}.
\label{torsion eq}
\end{equation}
where $C^{\mu}_{\;\;ab}\equiv e_a^{\ \nu}e_b^{\ \rho}C^{\mu}_{\;\;\nu\rho}$, and $C_a\equiv C^{\mu}_{\;\;a\mu}$. The {\it l.h.s.} of Eq.(\ref{torsion eq}) is, as expected, represented in terms of torsion. The source of torsion is called spin density.

Note that, contrary to curvature, torsion does not propagate around matter; because torsion has 24 components and Eq.(\ref{torsion eq}) consists of 24 equations, $C^{\mu}_{\;\;ab}=0$ is a unique solution if matter does not exist or if the Lagrangian of matter does not depend on $\omega^{ab}_\mu$
\footnote{On the subject of propagational torsion, for example, see Refs. \cite{belyaev,shapiro}.}. For example, in a system described by the Lagrangian of scalar or Yang-Mills fields, the torsion and the spin density vanish identically. Then the Einstein equation (\ref{E eq2}) returns to the usual one. This is of cause a result of the fact that gravitational theory without spinors does not need the spin connection. The usual Einstein equation is adequate enough to deal with the system of integer spin.

\section{General Relativity with torsion}

\subsection{Some modified formulae in General Relativity}

The existence of torsion alters some points in General Relativity, since Einstein's view point is to suppose $\Gamma^\rho_{\mu\nu}$ is symmetric. 
First, let us calculate the covariant divergence of asymmetric tensors $G_{\mu\nu}$ and $T_{\mu\nu}$. The previous definition of curvature is rewritten as
$[D_\mu, D_\nu]V^a=R^{ab}_{\ \ \mu\nu}V_b$,
which corresponds in the world components to
\begin{equation}
[\nabla_\mu, \nabla_\nu]V^\rho=
R^\rho_{\ \sigma\mu\nu}V^\sigma+C^{\sigma}_{\;\;\mu\nu}\nabla_\sigma V^\rho,
\label{riemann}
\end{equation}
where $R^\rho_{\ \sigma\mu\nu}$ is already given by Eq.(\ref{curvature}).
The Bianchi identity following from Eq.(\ref{riemann}) yields the covariant divergence of the Einstein tensor
\footnote{The reason for the contraction on the right side of indices is due to our notation, $R_{\mu\nu}\equiv e_{a\nu}R_\mu^{\;\; a}
=e_{a\nu}e_b^{\ \rho}R^{ba}_{\ \ \rho\mu}$.}:
\begin{equation}
\nabla^{\mu}G_{\nu\mu}=
C^{\rho\sigma\lambda}\left(\frac{1}{2}R_{\lambda\sigma\rho\nu}-R_{\rho\lambda}g_{\sigma\nu}\right).
\label{div G}
\end{equation}
Eq.(\ref{div G}) implies that, through the Einstein equation (\ref{E eq2}), the existence of torsion forces the usual covariant conservation law of the energy momentum tensor to be revised. Let us actually verify how this is done in the {\it r.h.s.} of Eq.(\ref{E eq2}).

In usual way, the symmetric energy momentum tensor of matter $\tau_{(\mu\nu)}$ is defined by varying the action of matter with respect to $g_{\mu\nu}$.
If torsion vanishes, because of the invariance of the matter action under a translation $x^\mu \rightarrow x^\mu +\xi^\mu(x)$, the covariant conservation law $\tilde{\nabla}^{\mu}\tau_{(\nu\mu)}=0$ holds, where $\tilde{\nabla}_\mu$ is the covariant derivative based on the Christoffel symbol.

In the case which torsion is nonvanishing, we should carefully examine the invariance of the action\cite{hehl,hehl2,utiyama}. Let us first consider the system described by world tensors without the vierbein and the spin connection for comparison with the vierbein formalism; let $\mathfrak{L}_{\rm m}=\mathfrak{L}_{\rm m}(g,\phi,\nabla\phi)$. According to Noether's procedure, when the equation of motion for $\phi$ is fulfilled, the invariance of the action under $x^\mu \rightarrow x^\mu +\xi^\mu(x)$ with the Lie derivatives
\begin{eqnarray}
\delta_{\rm L}g^{\mu\nu}&=&
2\left(
\nabla^{(\mu}\xi^{\nu)}+\xi^{\sigma}g^{\rho (\mu}C^{\nu)}_{\;\;\rho\sigma}
\right),
\label{GR lie derivative}
\\
\delta_{\rm L}\Gamma^\rho_{\nu\mu}&=&
-\nabla_\mu\nabla_\nu\xi^\rho-R^\rho_{\ \nu\sigma\mu}\xi^\sigma
+\nabla_\mu(C^{\rho}_{\;\;\sigma\nu}\xi^\sigma),
\end{eqnarray}
yields
\begin{equation}
\bar{\nabla}^{\mu}(\tau_{(\nu\mu)}+\bar{\nabla}_\rho s^\rho_{\ \mu\nu})=
C^{\mu\rho}_{\ \ \> \nu}(\tau_{(\mu\rho)}
+\bar{\nabla}_\sigma s^\sigma_{\ \rho\mu})
+s^\mu_{\ \rho\sigma}R^{\sigma\rho}_{\ \ \>\mu\nu},
\label{GR em conservation}
\end{equation}
where $\bar{\nabla}_{\mu}\equiv\nabla_\mu-C_\mu$ and
$
s^{\mu\nu}_{\ \ \> \rho}\equiv
(\partial\mathcal{L}_{\rm m}/\partial\nabla_\mu\phi^A)
(s^\nu_{\ \rho})^A_{\ B}\phi^B.
$

Thus we may conjecture that $\bar{\nabla}_\rho s^\rho_{\ \mu\nu}$ is responsible for the antisymmetric part of $\tau_{\mu\nu}$, although it is not yet clear whether $s^\rho_{\ \mu\nu}$ physically means the spin density or not.
Note that, if torsion does not exist, $s^\rho_{\;[\mu\nu]}$ vanishes identically and then the usual conservation law is recovered from Eq.(\ref{GR em conservation}). To observe this more manifestly, performing the partial integration of $\tau_{(\mu\nu)}$ in the change of the action with the Lie derivative (\ref{GR lie derivative}), we find
$
\bar{\nabla}^\mu\tau_{(\nu\mu)}=C^{(\mu\rho)}_{\ \ \ \ \nu}\tau_{(\mu\rho)}.
$

However, even if torsion exists, one can easily prove that $\tilde{\nabla}_{\mu}\tau^{(\nu\mu)}=0$ always holds in itself by using the definition of $\nabla_\mu$ and $\tilde{\nabla}_\mu$. This means that the existence of torsion does not alter the usual form of the conservation law.

Next, we consider the invariance of the action in the vierbein formalism. In this case, we have to establish the alternative Lie derivative with the local Lorentz transformation in mind. Our strategy to do this is to distinguish the local Lorentz transformation from the general coordinate transformation. On this subject, although Hehl {\it et.al.}\cite{hehl} organized the ``active" and ``passive" local Poincar\'e transformation by fixing the tetrad frame or the fields themselves, we wish to redefine the Lie derivative corresponding to their work.

Let us require the matter action to be invariant under both the general coordinate transformation $x^\mu\rightarrow x^\mu +\xi^\mu(x)$ and the Lorentz rotation $X^a\rightarrow X^a+\theta^a_{\ b}(x)X^b$. For this combination, any fields change as
\begin{equation}
\phi^A\rightarrow\phi^A+f^{\nu(A)}_\mu\partial_\nu\xi^\mu+\frac{i}{2}\theta^{ab}(S_{ab})^A_{\ B}\phi^B,
\end{equation}
where $f^{\nu(A)}_\mu$ denotes the generic coefficient of $\partial_\mu\xi^\nu$. Thus we define the combined Lie derivative:
\begin{equation}
\delta^*_{\rm L}\phi^A\equiv
f^{\nu(A)}_\mu\partial_\nu\xi^\mu+\frac{i}{2}\theta^{ab}(S_{ab})^A_{\ B}\phi^B
-\xi^\mu\partial_\mu\phi^A.
\label{comb.lie}
\end{equation}
In particular, Eq.(\ref{comb.lie}) applies to the vierbein as
\begin{equation}
\delta^*_{\rm L}e_a^{\ \mu}=
e_a^{\ \nu}\partial_\nu\xi^\mu+\theta_a^{\ b}e_b^{\ \mu}
-\xi^\nu\partial_\nu e_a^{\ \mu},
\label{vier lie derivative}
\end{equation}
which turns out to be consistent with the Lie derivative of the metric (\ref{GR lie derivative}) via Eq.(\ref{e square}) and the antisymmetricity of $\theta^{ab}$, i.e., $\delta^*_{\rm L}(e_a^{\ \mu}e^{a\nu})=\delta_{\rm L}g^{\mu\nu}$. As for the spin connection, since the Lorentz transformation law is given by Eq.(\ref{inf.tr.}), we have
\begin{equation}
\delta^*_{\rm L}\omega^{ab}_\mu=
-\omega^{ab}_\nu\partial_\mu\xi^\nu
+\theta^a_{\ c}\omega^{cb}_\mu+\theta^b_{\ c}\omega^{ac}_\mu
-\partial_\mu\theta^{ab}-\xi^\nu\partial_\nu\omega^{ab}_\mu.
\label{omega lie derivative}
\end{equation}
Hence, in the same manner as the previous case, but replacing $\delta_{\rm L}$ by $\delta_{\rm L}^*$, we can show that the identities following on the invariance of the action provide the conservation law including torsion:
\begin{eqnarray}
&&\!
(\mathcal{D}_\mu-C_\mu)S^\mu_{\;\;ab}=T_{[ab]},
\label{angular}
\\
&&\!\!\!\!\!\!\!\!\!\!\!
\bar{\nabla}^\mu T_{\nu\mu}=
C^{\mu\rho}_{\ \ \>\nu}T_{\mu\rho}+S^\mu_{\;\;ab}R^{ab}_{\ \ \mu\nu}.
\label{div T}
\end{eqnarray}
These are just covariant extensions of the conservation law\cite{duan, hammond} of the total angular momentum and the energy momentum in the Minkowski space-time, i.e., in the flat-limit, Eq.(\ref{angular}) supplies
$\partial_c(X_{[a}T_{b]}^{\ c}+S^c_{\;\;ab})=0$ together with Eq.(\ref{div T}),
$\partial_b T_a^{\ b}=0$. We also stress that Eq.(\ref{div G}) and Eq.(\ref{div T}) are compatible through the field equations (\ref{E eq2}) and (\ref{torsion eq}). That is why the definition of $\delta^*_{\rm L}$ is plausible and complementary to Hehl {\it et.al.}'s work.
It was thus that the revised covariant divergence of the energy momentum corresponding to Eq.(\ref{div G}) would be derived.

Note that, simultaneously with obtaining the above conservation law, the total derivative term in the change of the matter action yields two identities:
\begin{eqnarray}
&&\!\!\!\!\!\!\!\!\!\!
e(T_\mu^{\ \nu}-\Sigma_\mu^{\ \nu})+\partial_\rho U^{\rho\nu}_{\ \ \>\mu}=0,\\
&&\quad \ \ \ 
U^{(\mu\nu)}_{\ \ \ \ \rho}=0,
\end{eqnarray}
where $\Sigma_\mu^{\ \nu}$ represents the covariantized canonical energy momentum tensor:
\begin{equation}
\Sigma_\mu^{\ \nu}\equiv
\frac{\partial\mathcal{L}_{\rm m}}{\partial D_\nu\phi^A}D_\mu\phi^A
-\delta^\nu_\mu\mathcal{L}_{\rm m},
\label{canonical em}
\end{equation}
and $U^{\rho\nu}_{\ \ \>\mu}$ is defined by
$
U^{\rho\nu}_{\ \ \>\mu}\equiv
e(\partial\mathcal{L}_{\rm m}/\partial D_\rho\phi^A)f^{\nu(A)}_\mu
$
which allows the identification of $T_\mu^{\ \nu}$ with $\Sigma_\mu^{\ \nu}$ when $f^{\nu(A)}_\mu$ vanishes. For example, this identification is possible for Dirac fields which are world scalar, whereas it is impossible for Rarita-Schwinger fields because of their vectorial property.

Now, we consider splitting $G_{\mu\nu}$ into two parts by using Eq.(\ref{gamma}); one is constructed by the Christoffel symbol, and another results from torsion\cite{hehl}:
$
G_{\mu\nu}=\tilde{G}_{\mu\nu}+G_{\mu\nu}^{\rm (t)}.
$
Separating the Ricci tensor into the torsion-free part and the torsion part\cite{shapiro}, we find the expression of $G_{\mu\nu}^{\rm (t)}$:
\begin{eqnarray}
G_{\mu\nu}^{\rm (t)}&=&
\bar{\nabla}_\rho K^{\rho}_{\;\;\nu\mu}
-\nabla_\mu C_\nu+K^{\rho}_{\;\;\mu\sigma}K^{\sigma}_{\;\;\nu\rho}
\nonumber \\
&&{}
-\frac{1}{2}g_{\mu\nu}
\left[
-2\nabla_\rho C^\rho+C^\rho C_\rho
+g^{\lambda\tau}K^{\rho}_{\;\;\lambda\sigma}K^{\sigma}_{\;\;\tau\rho}
\right].
\label{Gt}
\end{eqnarray}
According to the field equation (\ref{torsion eq}) which torsion follows, one may replace torsion with the spin density identically. In spite of the existence of torsion, if $\tilde{G}_{\mu\nu}$ is regarded as a suitable quantity which describes gravitational theory all the same, Eq.(\ref{Gt}) can then be interpreted as a formula to provide the extra term related to spin for the energy momentum, i.e.,
\begin{equation}
\tilde{G}_{\mu\nu}=8\pi G\>T_{\mu\nu}-G_{\mu\nu}^{\rm (t)}
\equiv 8\pi G(\tilde{T}_{(\mu\nu)}+T_{\mu\nu}^{\rm (spin)}).
\label{spin correction def.}
\end{equation}
We should not forget that $T_{\mu\nu}$ can be decomposed into the torsion-free part and the torsion part by Eq.(\ref{omega}):
$T_{\mu\nu}=\tilde{T}_{\mu\nu}+T^{\rm (t)}_{\mu\nu}$.
$T_{\mu\nu}^{\rm (spin)}$ is automatically symmetric, and given by
\begin{eqnarray}
T_{\mu\nu}^{\rm (spin)}&=&
2\bar{\nabla}_\rho S_{(\mu\nu)}^{\ \ \ \>\rho}
-8\pi G \Bigg[
4S^{(\rho\sigma)}_{\ \ \ \ \mu}S_{(\rho\sigma)\nu}
-S_\mu^{\ \rho\sigma}S_{\nu\rho\sigma}-2S_\mu S_\nu
\nonumber \\
&&{}-\frac{1}{2}g_{\mu\nu}
\left(
4S^{(\rho\sigma)\lambda}S_{(\rho\sigma)\lambda}
-S^{\lambda\rho\sigma}S_{\lambda\rho\sigma}-2S^\rho S_\rho
\right)\Bigg]
+T^{\rm (t)}_{(\mu\nu)},
\label{full spin corr.}
\end{eqnarray}
with $S_\mu\equiv S^\nu_{\ \mu\nu}=C_\mu /(8\pi G)$. 
Eq.(\ref{full spin corr.}) clearly shows that there are extra contributions to both energy density and pressure of matter, if hydrodynamical description can be taken into account, through the products of the spin density which survive an averaging procedure\cite{nurgaliev}.
Of cause the reason to believe that $\tilde{G}_{\mu\nu}$ gives the essence of General Relativity is the fact $\tilde{\nabla}_\mu\tilde{G}^{\mu\nu}=0$.
Note that the antisymmetry $K_{\mu\nu\rho}=-K_{\nu\mu\rho}$ allows the metricity condition (\ref{metr.cond.}) to form $\tilde{\nabla}_\rho g_{\mu\nu}=0$\cite{penrose}. Therefore the substitution of the spin for torsion makes the geometry look like Riemannian.

Although the quantities in brackets in Eq.(\ref{full spin corr.}) are quite the same as the well-known correction\cite{hehl}, we regard Eq.(\ref{full spin corr.}) including the derivative of the spin density (and torsion part of $T_{\mu\nu}$) as the total correction, because $T_{(\mu\nu)}$ which is the symmetric part of the world component of $T_\mu^{\ a}$ originally differs from $\tau_{(\mu\nu)}$.
Provided that one can identify $T_{\mu\nu}$ with $\Sigma_{\mu\nu}$, the {\it r.h.s.} of Eq.(\ref{spin correction def.}) can be expressed in terms of $\tau_{(\mu\nu)}$ by the usual symmetrization procedure of the canonical energy momentum tensor.
We show some applications of the extended formulae (\ref{spin correction def.}) and (\ref{full spin corr.}) in the next subsection.

\subsection{Contribution of spinors to the Einstein equation}

Let us now give a specific example. We first consider the Dirac fields\cite{hehl,hehl2,shapiro,brill}. The Lagrangian of the Dirac field in a covariantized form is
\begin{equation}
\mathcal{L}_{\rm D}(e,\psi,D\psi)=
\frac{i}{2}\left[\>
\bar{\psi}\gamma^\mu D_\mu\psi-(\overline{D_\mu\psi})\gamma^\mu\psi
\>\right]-m\bar{\psi}\psi,
\label{dirac lagrangian}
\end{equation}
where $\overline{D_\mu\psi}\equiv\partial_\mu\bar{\psi}+(i/4)\omega^{ab}_\mu\bar{\psi}\bar{\sigma}_{ab}$ is the Dirac conjugation of Eq.(\ref{cov.spinor}) and $\gamma^\mu (x)\equiv e_a^{\ \mu}\gamma^a$ is a world component of the constant $\gamma$-matrices $\gamma^a$. The energy momentum tensor for the Dirac field is manifestly asymmetric, $T_{\mu\nu}\neq T_{\nu\mu}$. This means that torsion exists. In fact, the non-trivial spin density
\begin{equation}
S^\mu_{\;\;ab}=\frac{1}{8}\bar{\psi}\{\gamma^\mu,\sigma_{ab}\}\psi,
\end{equation}
leads to
$
C_{abc}=2K_{abc}=-16\pi G S_{abc}
$
by Eq.(\ref{torsion eq}), although its contraction $C_\mu$ vanishes because of the complete antisymmetry of $S_{abc}$.
As for the equation of motion, from Eq.(\ref{eom}) with $C_{\mu}=0$, we have
\begin{equation}
i\gamma^\mu D_\mu\psi-m\psi=0.
\label{dirac eq}
\end{equation}
The Lagrangian (\ref{dirac lagrangian}) vanishes for $\psi$ and $\bar{\psi}$ satisfying the equation of motion (\ref{dirac eq}), so that we find
\begin{equation}
T_{\mu\nu}=\frac{i}{2}\left[\>
\bar{\psi}\gamma_\nu D_\mu\psi-(\overline{D_\mu\psi})\gamma_\nu\psi
\>\right],
\label{dirac em2}
\end{equation}
which can be identified with the canonical energy momentum tensor (\ref{canonical em}).

Here we can derive the concrete expression of $T_{\mu\nu}^{\rm (spin)}$; we get
\begin{equation}
\tilde{G}_{\mu\nu}=8\pi G\>\tilde{\tau}_{(\mu\nu)}
+\frac{3}{16}g_{\mu\nu}(8\pi G\>\bar{\psi}\gamma_5\gamma_a\psi)^2,
\label{spin correction}
\end{equation}
where the formula in four dimensional space-time
$\{\gamma_c,\sigma_{ab}\}=2i\gamma_{abc}=-2\epsilon_{abcd}\gamma_5\gamma^d$
is employed.
The second term of the {\it r.h.s.} in Eq.(\ref{spin correction}) which appeared as a form of four-Fermi interaction is just the correction due to spin. Note that the only correction term emerged is proportional to $g_{\mu\nu}$ and hence may be expected to play the role of the effective cosmological constant.

Let us next apply Eq.(\ref{spin correction}) to the Robertson-Walker universe.
Taking a comoving frame $e_\mu^{\ 0}=u_\mu=(1,0,0,0)$, $e_\mu^{\ m}=0$ for $m=1,2,3$, and regarding $\tilde{\tau}_{(\mu\nu)}$ as perfect fluid filling in the universe, one finds the modified energy momentum tensor in the hydrodynamical description:
\begin{equation}
\tilde{\tau}_{(\mu\nu)}+T_{\mu\nu}^{\rm {(spin)}}
=(\rho+p)u_\mu u_\nu-(p-2\pi G\sigma^2)g_{\mu\nu},
\end{equation}
where $\sigma^2\equiv (3/4)(\bar{\psi}\gamma_5\gamma_a\psi)^2$, $\rho$ and $p$ indicate the energy density and the pressure of the Dirac field respectively, which are irrelevant to torsion. Thus total energy density and pressure are given by
\begin{eqnarray}
\rho_{\rm tot}&=&\rho+2\pi G\sigma^2,
\label{eff energy}
\\
p_{\rm tot}&=&p-2\pi G\sigma^2.
\end{eqnarray}
It is worth mentioning that the sign of the correction term in Eq.(\ref{eff energy}) is positive contrary to the existing semi-classical model\cite{gasperini,szydlowski}. In our approach, such a strange accident as ``the energy density of spin" becomes negative does not happen.

For the sake of simplicity, suppose that the universe is spatially flat. The scale factor $a(t)$ then evolves according to
\begin{equation}
\frac{\ddot{a}}{a}=-\frac{4\pi G}{3}(\rho+3p-4\pi G\sigma^2),
\end{equation}
which implies the accelerated expansion of the universe occurs provided that 
\begin{equation}
\rho+3p<4\pi G\sigma^2.
\label{inflation condition}
\end{equation}
It is quite possible to satisfy the condition (\ref{inflation condition}) in the early universe, by repeating the arguments which was done in the early work\cite{nurgaliev,gasperini}. We have to stress here that we need no thermodynamical variables assumed there, which have been introduced in order to describe the spinning fluid\cite{ray}, and to break the symmetric energy momentum tensor $\tau_{(\mu\nu)}$ of the fluid into the perfect fluid part and the spin part which is so crucial as to induce inflation\cite{gasperini}. Nevertheless, we can just split $\tau_{(\mu\nu)}$ into the perfect fluid part $\tilde{\tau}_{(\mu\nu)}$ which is irrelevant to torsion and the remaining spin part by separating torsion from the spin connection (\ref{omega}) and prove that inflation possibly occurs.

Next example is the Rarita-Schwinger fields which are essential objects in theory of supergravity\cite{deser,mitsuo,mitsuo2}. However, in the case where one considers the spin-3/2 field as mere matter, but not necessarily a supersymmetric partner of graviton, it is enough to treat the non-self-charge-conjugate field. Thus we start with the Rarita-Schwinger Lagrangian
\begin{equation}
\mathcal{L}_{3/2}(e,\psi,D\psi)=
-\frac{1}{2}\epsilon^{\mu\nu\rho\sigma}
\left[
\bar{\psi}_\mu\gamma_5\gamma_\nu D_\rho\psi_\sigma
-(\overline{D_\rho\psi_\mu})\gamma_5\gamma_\nu\psi_\sigma
\right]
-im_{\rm g}\bar{\psi}_\mu\sigma^{\mu\nu}\psi_\nu,
\label{RS lagrangian}
\end{equation}
where
$\epsilon^{\mu\nu\rho\sigma}\equiv
e_a^{\ \mu}e_b^{\ \nu}e_c^{\ \rho}e_d^{\ \sigma}\epsilon^{abcd}$
takes value of $\pm e^{-1}$ or 0.

The equation of motion (\ref{eom}) for $\psi_\mu$ is
\begin{equation}
\epsilon^{\mu\nu\rho\sigma}\gamma_5
\left(
\gamma_\nu D_\rho+\frac{1}{4}C^{\lambda}_{\;\;\nu\rho}\gamma_\lambda
\right)
\psi_\sigma+im_{\rm g}\sigma^{\mu\nu}\psi_\nu=0.
\label{RS eom}
\end{equation}
Similarly to the Dirac field, the Lagrangian (\ref{RS lagrangian}) vanishes for $\psi_\mu$ and $\bar{\psi}_\mu$ which fulfill the equation of motion (\ref{RS eom}). The energy momentum tensor then leads to
\begin{eqnarray}
T_\mu^{\ a}&=&\frac{1}{2}e_\nu^{\ a}\epsilon^{\lambda\nu\rho\sigma}
\left[
\bar{\psi}_\lambda\gamma_5\gamma_\mu D_\rho\psi_\sigma
-(\overline{D_\rho\psi_\lambda})\gamma_5\gamma_\mu\psi_\sigma
\right]
\nonumber \\
&&{}
-im_{\rm g}
(\bar{\psi}_\mu\sigma^{a\nu}\psi_\nu+\bar{\psi}_\nu\sigma^{\nu a}\psi_\mu
-e_\mu^{\ a}\bar{\psi}_\rho\sigma^{\rho\nu}\psi_\nu).
\label{RS em}
\end{eqnarray}
Note that Eq.(\ref{RS em}) can not be identified with the canonical energy momentum tensor, so that the covariant derivative of the spin density has to be taken into account in Eq.(\ref{full spin corr.}).
In fact, contrary to the Dirac field, the spin density is not completely antisymmetric for the Rarita-Schwinger field and reads
\begin{equation}
S_{abc}=\frac{i}{4}
\left[
\bar{\psi}_b\gamma_a\psi_c-\bar{\psi}_c\gamma_a\psi_b
+\eta_{ab}(\bar{\psi}_c\gamma^d\psi_d-\bar{\psi}_d\gamma^d\psi_c)
-\eta_{ac}(\bar{\psi}_b\gamma^d\psi_d-\bar{\psi}_d\gamma^d\psi_b)
\right],
\end{equation}
which causes nonvanishing torsion,
$C_{abc}=-4\pi Gi(\bar{\psi}_b\gamma_a\psi_c-\bar{\psi}_c\gamma_a\psi_b)$.

Now we can derive the spin correction (\ref{full spin corr.}) for the Rarita-Schwinger field:
\begin{eqnarray}
T_{\mu\nu}^{\rm (spin)}&=&
2\tilde{\nabla}_\rho S_{(\mu\nu)}^{\ \ \ \>\rho}
+8\pi G \Bigg[
4J^{(\rho\sigma)}_{\ \ \ \ \mu}J_{\nu(\rho\sigma)}
-2J_\sigma^{\ \rho}{}_{\!(\mu}J^\sigma_{\;\;\nu)\rho}
+4J^\rho J_{(\mu\nu)\rho}
\nonumber \\
&&{}+4J_\mu J_\nu
+\frac{1}{2}g_{\mu\nu}
\left(
J^{\rho\sigma\lambda}J_{\rho\sigma\lambda}
+2J^{\rho\sigma\lambda}J_{\sigma\rho\lambda}-4J^\rho J_\rho
\right)\Bigg],
\label{RS spin corr.}
\end{eqnarray}
where
$J_{\mu\nu\rho}\equiv
(\bar{\psi}_\mu\gamma_\nu\psi_\rho-\bar{\psi}_\rho\gamma_\nu\psi_\mu)/4$ and
$J_\mu\equiv J^\nu_{\ \nu\mu}$. Although the covariant derivative term based on the Christoffel connection is left over in Eq.(\ref{RS spin corr.}), this term is expected to vanish by the averaging procedure\cite{nurgaliev}.

Eq.(\ref{RS spin corr.}) shows that there are three types of the contribution to the pressure. One renders the pressure positive, another acts inversely to push outside like the cosmological constant, and the other is indefinite which effects to have.

To observe this, let us define composite fields
\begin{eqnarray}
&&\!\!\!\!\!\!\!\!\!
J^{i\mu 0}J_{i\mu 0}\equiv \frac{\chi^2}{4},
\ J^{i\mu j}J_{i\mu j}\equiv \vartheta^2,
\ J^{ij0}J_{0ij}\equiv \frac{\zeta}{4},
\ J^{ikj}J_{ijk}\equiv \frac{\eta}{2},
\nonumber \\
&&\ \ J^{ij0}J_{i0j}\equiv \frac{\varphi}{6},
\ J^{00i}\equiv \frac{\tau^i}{2},
\ J^i_{\ i0}\equiv \frac{\lambda}{2\sqrt{2}},
\ J^j_{\ ji}\equiv \frac{\upsilon_i}{2},
\end{eqnarray}
where the indices $i,j,k$ run the spatial components.
Taking the comoving frame again, we have
\begin{eqnarray}
\rho_{\rm tot}+3p_{\rm tot}=
\rho+3p
+8\pi G
\left[
\lambda^2+(\tau_i+\upsilon_i)^2-(\chi^2+\vartheta^2)
+\zeta-\eta-\varphi+\tau_i\upsilon^i
\right].
\label{acc.cond.}
\end{eqnarray}
Therefore whether the universe gets accelerated or not depends on the balance of values of composite fields.
However, it is the terms related to $J_\mu$ that prevent the universe from accelerating.
So let us tentatively impose the irreducible condition $\gamma^a\psi_a=0$ in the Minkowski space-time on the expression (\ref{RS spin corr.}). Then, we find the condition of the acceleration relaxed:
\begin{eqnarray}
\rho_{\rm tot}+3p_{\rm tot}=
\rho+3p
+8\pi G
\left[
-(\chi^2+\vartheta^2+\tau_i^2)
+\zeta-\eta-\varphi
\right],
\end{eqnarray}
Note that, for comparison with Eq.(\ref{acc.cond.}), the irreducible condition not only eliminates the contribution making the pressure positive but also adds the negative term which raises the acceleration.

Thus it is not impossible that the Rarita-Schwinger field triggers off the acceleration of the universe.
However, the cosmological roles of the irreducible condition are not yet clear. We need further studies on this subject in order to construct the more realistic cosmological model and to explain the early or recent accelerated expansion of the universe\cite{tomoki}.
Finally note that, contrary to the Dirac field, the energy density of the Rarita-Schwinger spin is not necessarily positive.

\section{Conclusions}

We derived the some formulae modified by torsion in General Relativity.
The field equation obtained by taking the variation with respect to the spin connection solely determines torsion as a function of the spin density which is non-trivial for the Dirac and the Rarita-Schwinger fields.

In the presence of torsion, it was shown that either side of the Einstein equation has the nonvanishing covariant divergence. This fact is however self-consistent in the framework under consideration.
To derive this, we employed Noether's procedure with the Lie derivative for which the general coordinate transformation and the local Lorentz rotation are combined.

Subsequently, for the Dirac and the Rarita-Schwinger fields, the correction to the energy momentum tensor due to spin was explicitly calculated  without introducing thermodynamical variables assumed in the existing cosmological model. The results obtained imply that spinor fields can provide the negative pressure and be an alternative to false vacuum in the early stage of the universe.

Let us close with two remarks. First, gravitational theory with the massless Rarita-Schwinger action possesses the most remarkable symmetry, i.e., supersymmetry mixing the spin-2 graviton and the spin-3/2 gravitino\cite{nieuwen,nilles,weinberg,deser}. For the early evolution of the universe, one may hope to construct the plausible cosmological model in the context of supersymmetry\cite{mitsuo2}. Second, the spin density is so tiny in large scale that usual General Relativity and Cartan's theory are presently indistinguishable. However, as the cosmological constant has caused the accelerated expansion of the present universe, so the effect of the spin density might appear in the future\cite{tomoki}. These remarks will give us further tasks.
\newline

One of the authors (T. W.) wishes to express his thanks to I. Aizawa, M. Ishiguro, Y. Okame and T. Fukuda for very useful discussions, and also to Prof. Y. Kawamura for valuable comments at the summer institute at Yamanaka lake. M.J.H would like to express his sincere condolence to Prof. Tadao Nakano died on August 15th, 2004, whose  distinguished works and intuition on particle physics and general relativity continually have been  a source of his insights and motivations on physics and life.

\eject
\end{document}